\documentclass[lettersize,journal]{IEEEtran}
\usepackage{amsmath,amsfonts}
\usepackage{algorithmic}
\usepackage{array}
\usepackage[caption=false,font=normalsize,labelfont=sf,textfont=sf]{subfig}
\usepackage{textcomp}
\usepackage{stfloats}
\usepackage{url}
\usepackage{verbatim}
\usepackage{graphicx}
\usepackage{hyperref}
\usepackage{epstopdf}
\usepackage{booktabs}  
\usepackage{tabularx}  
\usepackage{ragged2e}  
\usepackage{makecell}  
\usepackage{comment}
\def\BibTeX{{\rm B\kern-.05em{\sc i\kern-.025em b}\kern-.08em
    T\kern-.1667em\lower.7ex\hbox{E}\kern-.125emX}}
\usepackage{balance}
\begin{document}
\title{Rethinking Security in Semantic Communication: Latent Manipulation as a New Threat}
\author{Zhiyuan Xi, Kun Zhu,~\IEEEmembership{Member,~IEEE}

\thanks{Z.Y. Xi and K. Zhu are with the College of Computer Science and Technology, Nanjing University of Aeronautics and Astronautics, Nanjing 210016, China (e-mail: \{xizhiyuan, zhukun\}@nuaa.edu.cn).}
}

\maketitle

\begin{abstract}
Deep learning–based semantic communication (SemCom) has emerged as a promising paradigm for next-generation wireless networks, offering superior transmission efficiency by extracting and conveying task-relevant semantic latent representations rather than raw data. However, the openness of the wireless medium and the intrinsic vulnerability of semantic latent representations expose such systems to previously unrecognized security risks. In this paper, we uncover a fundamental latent-space vulnerability that enables Man-in-the-Middle (MitM) attacker to covertly manipulate the transmitted semantics while preserving the statistical properties of the transmitted latent representations. We first present a Diffusion-based Re-encoding Attack (DiR), wherein the attacker employs a diffusion model to synthesize an attacker-designed semantic variant, and re-encodes it into a valid latent representation compatible with the SemCom decoder. 
Beyond this model-dependent pathway, we further propose a model-agnostic and training-free Test-Time Adaptation Latent Manipulation attack (TTA-LM), in which the attacker perturbs and steers the intercepted latent representation toward an attacker-specified semantic target by leveraging the gradient of a target loss function. In contrast to diffusion-based manipulation, TTA-LM does not rely on any generative model and does not impose modality-specific or task-specific assumptions, thereby enabling efficient and broadly applicable latent-space tampering across diverse SemCom architectures.
Extensive experiments on representative semantic communication architectures demonstrate that both attacks can significantly alter the decoded semantics while preserving natural latent-space distributions, making the attacks covert and difficult to detect. Our findings reveal a critical and previously overlooked security gap in latent-space transmission and underscore the urgent need for mechanisms that ensure semantic integrity and authentication in SemCom systems.
\end{abstract}


\section{Introduction}
Recent progress in artificial intelligence (AI) has led to increasing interest in communication systems that convey meaning rather than raw data. While classical communication theory focuses on the reliable transmission of symbol sequences, many current applications~\cite{uav_1, uav_2, uav_3} require the exchange of task-relevant information that reflects the underlying intent of the data. This requirement has motivated the development of semantic communication (SemCom)~\cite{survey_1}, where semantic latent representations are extracted and transmitted to support more efficient and effective communication.

Recent works have proposed a variety of deep learning–based SemCom systems for image, speech, text, and multimodal data. These systems typically adopt learned semantic encoders, such as VQ-VAE~\cite{vqvae_sc} or transformer models~\cite{swinjscc}, to map data into compact latent representations that capture the essential information needed for downstream tasks. Transmitting these semantic latent representations can reduce bandwidth usage and improve robustness to channel noise, while preserving task performance.

\begin{figure}[htbp]
    \centering
    \noindent\includegraphics[width=3.5in]{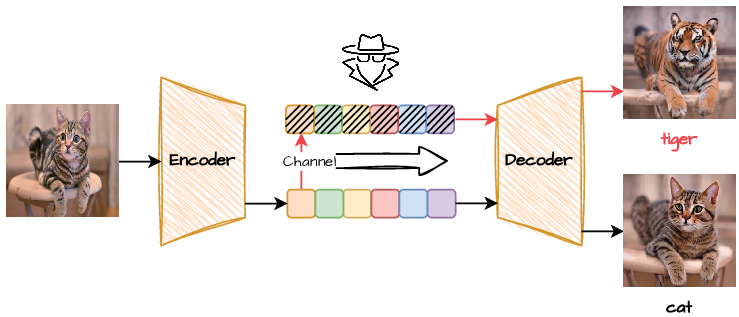}
    \caption{Latent-Space Semantic Manipulation by an MitM Attacker in a SemCom System}
    \label{tto_attack_introduction}
\end{figure}

However, the security implications of semantic communication~\cite{secure_sc} remain insufficiently explored. Existing research primarily focuses on physical-layer security, adversarial robustness at the input level, and semantic noise mitigation. In practical deployments, SemCom systems typically rely on a standardized semantic encoder–decoder pair across devices, analogous to conventional communication systems where coding and modulation schemes are uniformly adopted. Consequently, all users operate with an identical latent-space structure and decoding process. Since these latent representations carry structured, task-critical semantic information, even small or imperceptible perturbations can induce substantial deviations in the reconstructed output. Nevertheless, current SemCom frameworks lack mechanisms to verify the integrity or authenticity of transmitted latents, leaving the semantic reconstruction process exposed to tampering. As illustrated in Fig. ~\ref{tto_attack_introduction}, a MitM attacker can tamper with the transmitted latent representation so that an input image of a cat is maliciously decoded as a tiger instead of the original cat.

\begin{figure}[htbp]
    \centering
    \noindent\includegraphics[width=3.5in]{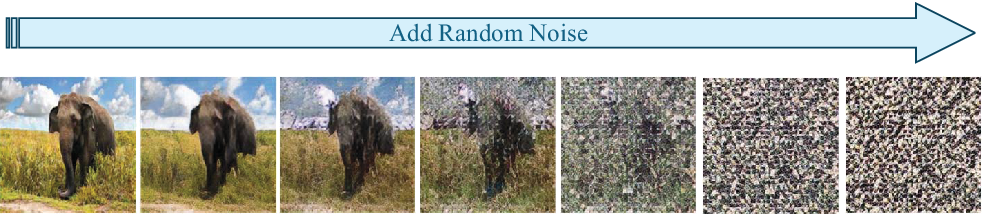}
    \caption{The reconstruction results decoded by noise-injected latent representations}
    \label{noise_example}
\end{figure}

In practice, manipulating a semantic latent representation in a controlled manner is highly challenging. A naive strategy, such as directly injecting random perturbations or noise into the latent representations, typically does not yield meaningful semantic changes. For reconstruction-oriented SemCom systems, such perturbations merely degrade the latent structure learned by the encoder, causing the decoder to produce outputs that are blurred, distorted, or heavily contaminated by noise as shown in Fig.~\ref{noise_example}. These artifacts are easily detectable and do not resemble natural semantic deviations. 
Therefore, the perceived difficulty of deliberately manipulating latent representations has led existing research to largely overlook this sort of attack. 

In this work, however, we demonstrate that semantic latent representations can in fact be efficiently and reliably tampered with, even without relying on any auxiliary generative models. To demonstrate this, we design two complementary latent-space attack mechanisms. The first, a Diffusion-based Re-encoding Attack (DiR), employs a latent diffusion model to synthesize attacker-specified semantics and re-encode them into valid latent representations compatible with the semantic decoder. The second, a Test-Time Adaptation Latent Manipulation (TTA-LM) attack, perturbs the intercepted latent representation toward an attacker-defined semantic objective in a training-free and model-agnostic manner, thereby illustrating that effective latent-space tampering can be achieved even without additional models.

The main contributions of this work are summarized as follows:

1) \textbf{Latent-space vulnerability of semantic communication.}
We identify and formalize a previously overlooked attack in deep learning–based semantic communication systems, showing that the transmitted semantic latent representations can be covertly tampered with by a MitM attacker while remaining statistically consistent with those produced by the legitimate encoder.

2) \textbf{Diffusion-based re-encoding attack.}
We propose a Diffusion-based Re-encoding Attack (DiR) that leverages a latent diffusion model to synthesize attacker-specified semantic variants and re-encode them into valid latent representations compatible with the shared semantic decoder, thereby enabling targeted semantic alteration during transmission.

3) \textbf{Training-free and model-agnostic latent manipulation.}
We develop a Test-Time Adaptation Latent Manipulation (TTA-LM) attack that perturbs intercepted latent representations toward attacker-defined semantic objectives using gradient-based adaptation. TTA-LM operates in a training-free and model-agnostic manner, demonstrating that effective latent-space tampering is possible even without any auxiliary generative models.

4) \textbf{Comprehensive empirical evaluation and security insights.}
We conduct extensive experiments on representative semantic communication architectures, showing that both attacks can substantially change the decoded semantics while preserving natural latent-space statistics, making them difficult to detect. We further discuss the security implications of these findings and highlight the need for semantic integrity and authentication mechanisms tailored to latent-space transmission.

\section{Related Work}
\subsection{Deep Learning-based Semantic Communications}

Deep learning has played a central role in advancing semantic communication systems. Unlike conventional communication, which emphasizes symbol-level accuracy, deep learning-based semantic communication (SemCom) seeks to extract task-relevant semantic features and transmit them in a compact form. A typical SemCom system consists of a semantic encoder at the transmitter and a semantic decoder at the receiver, with both sides optionally maintaining auxiliary knowledge to support semantic understanding.

A wide range of architectures have been developed under this paradigm. In the vision domain, DeepSC-RI~\cite{vit_sc} introduces a dual-branch multi-scale semantic extractor and a cross-attention fusion module to enhance robustness under diverse channel conditions and impairment levels. Chen \textit{et al.}~\cite{RVQ_sc} present a low-bitrate digital semantic speech communication framework that combines RVQGAN-based multiscale semantic coding with a U-Net–based noise suppression module, enabling high-quality reconstruction in low-SNR environments.

In the speech domain, DeepSC-TS~\cite{speech_swin_sc} employs a Transformer-based design to reconstruct and integrate multi-level semantic information while suppressing semantic noise without increasing transmission overhead. Weng \textit{et al.}~\cite{speech_sc} further propose DeepSC-ST, which transmits recognition-related semantic features and performs speech synthesis at the receiver, significantly reducing transmitted data volume while maintaining task performance.

In the text domain, Peng \textit{et al.}~\cite{text_sc_1} introduce R-DeepSC, a robust text semantic communication system equipped with a semantic corrector to mitigate textual impairments, and later extend it with a non-autoregressive variant that achieves parallel semantic decoding with reduced complexity. Mao \textit{et al.}~\cite{text_sc_2} propose Ti-GSC, a GAN-assisted semantic communication framework that eliminates the need for channel state information by jointly learning semantic encoding/decoding and distortion suppression to counter syntactic and semantic degradation over fading channels.

While these advances significantly improve reconstruction fidelity, task performance, and robustness to channel impairments, the security of the transmitted latent representations has received comparatively limited attention. These latent features carry structured semantic information and directly determine the receiver's decoded output, making their integrity essential for reliable semantic communication.

\subsection{Security Threats in Semantic Communications}

Semantic communication introduces security risks that differ from those in conventional systems. Because semantic encoders transmit compact latent representations, disturbances that do not visibly alter the transmitted signal may still affect the meaning reconstructed at the receiver. Existing studies~\cite{security_survey_1, security_survey_2} highlight several major threat categories:
\begin{itemize}
\item \textbf{Adversarial perturbations and semantic tampering}: small modifications to inputs or transmitted features can lead to incorrect or misleading semantic outputs.
\item \textbf{Semantic eavesdropping}: intercepted latent representations may reveal high-level information when processed using powerful inference models.
\item \textbf{Model-level threats}: the reliance on deep learning exposes SemCom systems to model theft, poisoning, and computational attacks that degrade system availability.
\end{itemize}

\begin{table*}[t]
\centering
\caption{Security Threat Categories and Representative Works in Semantic Communications}
\label{tab:security_threats}
\renewcommand{\arraystretch}{1.3}
\newcolumntype{L}{>{\RaggedRight\arraybackslash}X}

\begin{tabularx}{\textwidth}{
    >{\hsize=0.6\hsize}L  
    >{\hsize=1.4\hsize}L  
    >{\hsize=1.0\hsize}L  
}
\toprule
\textbf{Threat Category} & \textbf{Description} & \textbf{Representative Works} \\
\midrule

Adversarial perturbations and semantic tampering 
& Crafted perturbations on inputs or transmitted features that mislead semantic decoding. 
& Sagduyu \textit{et al.}~\cite{secure_sc_1}; \newline
  Hoang \textit{et al.}~\cite{secure_sc_2}; \newline
  Hua \textit{et al.}~\cite{secure_sc_5}; \newline
  Anjum \textit{et al.}~\cite{secure_sc_6} \\
\addlinespace[1ex] 

Semantic eavesdropping 
& Inferring or reconstructing original content from intercepted semantic features.  
& Chen \textit{et al.}~\cite{secure_sc_7} \\
\addlinespace[1ex]

Model-level threats 
& Attacks on semantic encoders/decoders including model poisoning, backdoors, and computational degradation. 
& Zhou \textit{et al.}~\cite{secure_sc_3, secure_sc_4}; \newline
  Hua \textit{et al.}~\cite{secure_sc_5}; \newline
  Anjum \textit{et al.}~\cite{secure_sc_6} \\

\bottomrule
\end{tabularx}
\end{table*}

Recent work has further exposed the vulnerability of deep learning–based SemCom systems to adversarial manipulation across different modalities and attack surfaces. Sagduyu \textit{et al.}~\cite{secure_sc_1} show that semantic communication pipelines are susceptible to multi-domain perturbations applied either at the transmitter input or to the received channel signal, where even small, coordinated disturbances can significantly distort semantic outputs despite low reconstruction errors. Hoang \textit{et al.}~\cite{secure_sc_2} extend these findings to speech-oriented SemCom, demonstrating that channel-aware gradient-based perturbations, both targeted and non-targeted, can severely reduce semantic quality under AWGN, Rayleigh, and Rician channels, outperforming conventional jamming attacks.

Beyond perturbation-based threats, poisoning and backdoor attacks have also been explored. Zhou \textit{et al.}~\cite{secure_sc_3} introduce BASS, a backdoor paradigm capable of manipulating reconstructed semantic symbols through poisoned training samples, and propose defenses based on coordinated training, trigger estimation, and pruning. A follow-up work by Zhou \textit{et al.}~\cite{secure_sc_4} presents stealthier backdoors enabled by transformation-based target generation and latent-space trigger optimization, allowing backdoors to evade both visual and semantic detection while maintaining high attack success rates.

Other studies investigate modality-specific vulnerabilities. In the speech domain, Hua \textit{et al.}~\cite{secure_sc_5} demonstrate that DeepSC-based systems are susceptible to imperceptible and targeted physical-layer attacks that induce controlled transcription errors under realistic wireless channels while remaining nearly inaudible. For text-based SemCom, Anjum \textit{et al.}~\cite{secure_sc_6} propose SemPerGe, a black-box semantic perturbation framework that identifies influential tokens and generates linguistically natural substitutions, achieving high-success semantic drift without requiring access to model internals. In addition, Chen \textit{et al.}~\cite{secure_sc_7} uncover a fundamental privacy risk: transmitted semantic features can be inverted to reconstruct recognizable images, indicating that semantic latents retain substantial recoverable information even under channel noise.

Despite these advances, existing attacks primarily operate on raw inputs, physical waveforms, or poisoned training data, and therefore cannot directly manipulate the semantics carried by the transmitted latent representations. Their effects are largely limited to inducing misclassification or semantic drift and often rely on modality-specific assumptions or gradient access. As a result, the security of latent representations, central to modern semantic communication systems, remains insufficiently explored.

\subsection{Security Defenses in Semantic Communications}

Several recent efforts have explored defense mechanisms for enhancing the confidentiality and robustness of semantic communication systems. One line of work adopts physical-layer security (PLS) to limit semantic leakage during transmission, where the semantic encoder–decoder is jointly optimized to preserve semantic fidelity for legitimate users while suppressing recoverable information at eavesdroppers under secrecy constraints~\cite{secure_sc_pls}. Another line focuses on architectural obfuscation and black-box attack resistance: semantic-block coding, variable semantic coding, and hybrid-channel strategies are used to enlarge the semantic coding space and reduce the predictability of encoder–decoder mappings, making model extraction and query-based attacks more difficult~\cite{secure_sc_blackbox}. Generative models have also been incorporated into secure semantic pipelines, for example by using diffusion- or GAN-based modules to assist covert or distortion-suppressed semantic transmission~\cite{secure_sc_genai,secure_sc_img}.

In parallel, blockchain and cryptographic techniques have been introduced to protect semantic data in decentralized environments. Lin \textit{et al.} propose a blockchain-aided semantic communication framework for AIGC services in Metaverse and design a zero-knowledge–proof–based defense mechanism that records and verifies semantic transformations on-chain to distinguish adversarial semantic data from authentic ones~\cite{secure_sc_blockchain}. 

Although a variety of defense strategies have been explored, most of them still rely on channel advantages, coding randomization, or external trust infrastructures, and rarely address attacks that operate directly on the transmitted latent representations, leaving the core semantic features of modern SemCom systems insufficiently protected.

\subsection{Generative AI and Latent Space Manipulation}

Recent advances in generative AI have led to models that perform synthesis and editing directly in learned latent spaces. Latent diffusion models (LDMs), proposed by Rombach \textit{et al.}~\cite{ldm}, move the diffusion process from pixel space to a VAE-compressed latent space, substantially reducing computational cost while preserving high-level semantics and supporting flexible conditioning mechanisms. On the representation side, a highly compressed tokenizer~\cite{hc_tokenizer} has been introduced to map data into extremely compact codes that still retain sufficient semantic structure to support generation, demonstrating that powerful generative behavior can emerge from very low-dimensional latent tokens.

Building on these ideas, a feature-aligned tokenizer~\cite{recon_gen} is proposed to balance reconstruction fidelity and generative quality by imposing stronger structure on latent features, alleviating the optimization dilemma of latent diffusion models. LatentPaint~\cite{latentpaint} further shows that image inpainting and related editing tasks can be carried out entirely in latent space by conditioning the diffusion process on partial observations, enabling efficient, training-free semantic editing. Beyond diffusion, SDGAN~\cite{sdgan} disentangles facial attributes in the latent space of a GAN, allowing controllable and region-aware semantic manipulation. These works collectively indicate that latent representations are semantically rich and easily steerable, which motivates our investigation of the security risks arising from adversarial manipulation of transmitted latents in semantic communication systems.

\section{System Model}

\subsection{Semantic Communication model}
We consider a typical semantic communication system consisting of a sender, a receiver, and a wireless channel between them. Let $x$ denote the source data (e.g., an image or multi-modal sensory input). The sender employs a semantic encoder $f_{\theta}(\cdot)$, producing a compact latent representation:
$$z = f_{\theta}(x).$$

Unlike conventional communication systems that transmit raw signals such as pixels, semantic systems transmit only the latent representation $z$, which contains high-level semantic meaning. The receiver applies a semantic decoder $g_{\phi}(\cdot)$ to reconstruct the semantic content:
$$\hat{x} = g_{\phi}(z).$$ 

Following the design principle of conventional communication systems, where coding/decoding modules such as LDPC, Turbo codes, and modulation schemes are standardized and shared, semantic communication systems are expected to deploy a unified semantic encoder and decoder across all devices. Thus, \textbf{(a)} all communication parties use the same encoder–decoder architecture, \textbf{(b)} the latent space structure, codebook format, and dimensionality are publicly known, and \textbf{(c)} the encoder and decoder parameters are fixed after pre-training and remain unchanged during deployment.

\subsection{Channel Model}
The wireless channel between the sender and the receiver is modeled as a noisy channel. The transmitted latent representations $z$ perturbed by additive noise:
$$z_{c} = z + n,$$
where $n$ represents channel noise.

\subsection{Attacker Model}
We consider an in-path attacker who can observe and intercept the transmitted latent representation. Leveraging the publicly known codec specification, the attacker can modify or replace the latent code with a crafted representation $z'$. The attacker does not control or modify the encoder or decoder and only operates on the transmitted latent. 

The attacker aims to induce the receiver to reconstruct content aligned with an attacker-specified semantic target $x_t$. This requires generating a manipulated latent representation that satisfies: 
$$g_{\phi}(z') \approx x_t,$$
while ensuring that $z'$ remains statistically similar to legitimate latent codes so that the manipulation remains undetected.

\section{Diffusion-based Regenerate-and-Recode Attack}
\begin{figure*}[htbp]
    \centering
    \noindent\includegraphics[width=6in]{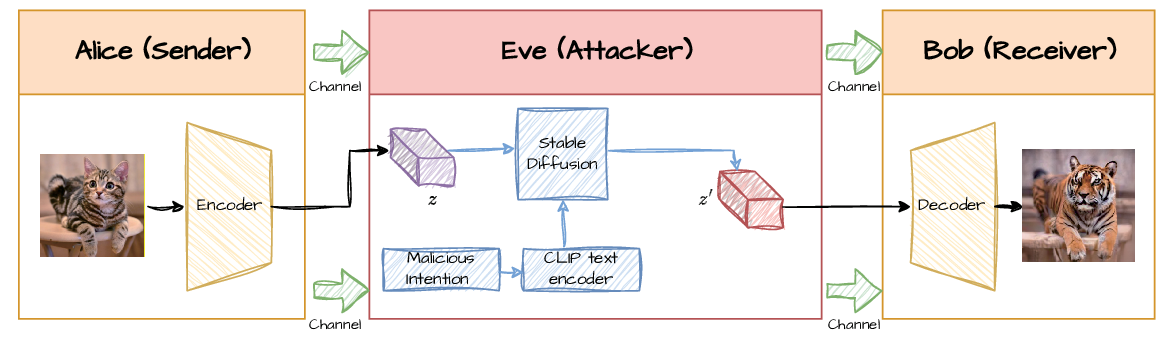}
    \caption{Diffusion-based Regenerate-and-Recode Attack}
    \label{DiR}
\end{figure*}

In this section, we introduce a Diffusion-based Re-encoding attack (DiR). Fig. ~\ref{DiR} shows the proposed DiR pipeline. After intercepting the transmitted latent representation $z$, the attacker directly performs semantic editing in the latent space using a latent diffusion model (LDM). This design eliminates the need for image-level reconstruction, making the attack more efficient and less detectable.

\subsection{Text-Conditioned Latent Diffusion Editing}
The attacker specifies a malicious semantic target through a text prompt $t$. A CLIP-based text encoder produces the conditioning embedding:
$$e_t = \mathrm{TextEnc}(t).$$
A latent diffusion model is then applied directly to the intercepted latent $z$. Following the standard DDPM formulation in latent space, the forward noising process is:
$$z_T = \sqrt{\bar{\alpha}_T}\, z + \sqrt{1 - \bar{\alpha}_T}\, \epsilon,
\qquad \epsilon \sim \mathcal{N}(0,I),$$
where $\alpha_t$ denotes the noise schedule.

During the reverse denoising process, the model iteratively predicts the noise conditioned on the attacker’s text embedding:
$$z_{t-1} 
= \frac{1}{\sqrt{\alpha_t}}
\left(
    z_t - (1 - \alpha_t)\,\epsilon_\psi(z_t, t, e_t)
\right)
+ \sigma_t \xi,
\quad \xi \sim \mathcal{N}(0,I),$$
where $\epsilon_\psi$ is the denoiser of the latent diffusion model.

After $T$ reverse steps, the attacker obtains the manipulated latent representation:
$$z' = D_{\mathrm{LDM}}(z, e_t),$$
which embeds the semantics of the target prompt of attackers while remaining in the same latent space as the encoder output.

\subsection{Latent Replacement}
Because LDM outputs latent codes that follow the same distribution as the original encoder latent, the attacker can directly replace the transmitted representation through $z'$. The receiver uses the shared semantic decoder to reconstruct:
$$\hat{x} = g_{\phi}(z').$$

\section{Test-Time Adaptation Latent Manipulation Attack}
\begin{figure*}[htbp]
    \centering
    \noindent\includegraphics[width=6in]{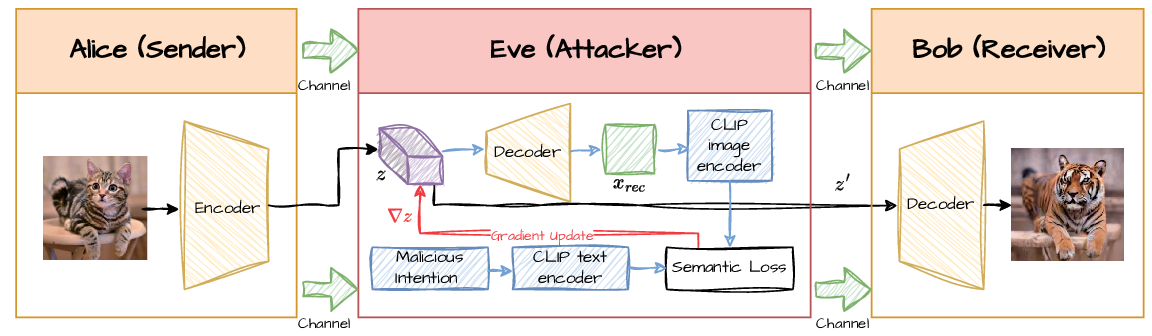}
    \caption{Test-Time Adaptation Latent Manipulation Attack}
    \label{TTA-LM}
\end{figure*}

In this section, we introduce a Test-Time Adaptation Latent Manipulation Attack (TTA-LM). Fig. ~\ref{TTA-LM} shows the proposed TTA-LM pipeline. Unlike diffusion-based approaches, TTA-LM does not rely on any external generative model. The attacker only performs gradient updates on the intercepted latent representation using the shared decoder and a CLIP-based semantic loss. This eliminates the need for a computationally expensive diffusion model, significantly reducing attack cost and enabling real-time manipulation with minimal resources. As a result, the TTA-LM is lightweight, requires no additional model deployment, and is much easier for an attacker to execute in practical semantic communication scenarios.

\subsection{Semantic Loss and Gradient Backpropagation}
After intercepting the transmitted latent representation $z$, the attacker first reconstructs an approximate image using the shared semantic decoder:
$$x_{rec} = g_{\phi}(z).$$
Although $x_{rec}$ may not be identical to the sender’s original image, it provides a differentiable semantic estimate that can be processed by a vision-language model.

Similar to DiR, the attacker specifies a malicious semantic target through a text prompt $t$. A CLIP text encoder generates the textual embedding:
$$e_t = \mathrm{TextEnc}(t).$$
Simultaneously, the reconstructed image is encoded by a CLIP image encoder:
$$e_x = \mathrm{ImgEnc}(x_{rec}).$$

To steer the latent representation toward the attacker’s target semantics, the attacker defines a CLIP-based semantic loss:
$$\mathcal{L}_{sem}
= 1 - \cos\left( e_x,\, e_t \right),$$
which encourages the reconstructed image’s embedding to align with the target textual embedding.

Crucially, we treat the intercepted latent representation $z$ as an optimizable parameter. Therefore, gradients $\nabla_z \mathcal{L}_{sem}$ of the semantic loss with respect to $z$ can be directly computed and used for iterative updates:
$$z^{(k+1)} = z^{(k)} - \eta \,\nabla_z \mathcal{L}_{sem},$$
where $\eta$ is a small step size. This adaptation updates only the latent representation, without modifying any model parameters, making the attack lightweight and efficient.

\subsection{Latent Replacement}
After convergence, the attacker obtains a manipulated latent representation:
$$z' = z^{(K)}.$$
The receiver directly decodes the manipulated latent:
$$\hat{x} = g_{\phi}(z'),$$
producing an output aligned with the attacker’s semantic target.

\section{Experiments}
\subsection{Experiments Settings}

\subsubsection{Baseline Semantic Encoder-Decoder}
For all experiments, the transmitter and receiver share an identical semantic encoder–decoder pair, following the common design of modern semantic communication systems.

We instantiate the semantic encoder–decoder with four representative model:

\noindent
\textbf{VQGAN}~\cite{vqgan}: a convolutional VQ-based tokenizer trained with reconstruction, perceptual, and adversarial losses to obtain high-fidelity discrete image codes suitable for transformer modeling.

\noindent
\textbf{TiTok}~\cite{titok}: a 1D image tokenizer that compresses each image into a very short sequence (e.g., 32) of discrete tokens while maintaining good reconstruction quality.

\noindent
\textbf{One-D-Piece}~\cite{one_d}: an extension of 1D tokenization that supports quality-controllable, variable-length token sequences, enabling a trade-off between bitrate and reconstruction fidelity.

\noindent
\textbf{IBQ}~\cite{ibq}: a scalable vector-quantization tokenizer that updates code indices via index backpropagation, achieving high codebook utilization and strong reconstruction performance with large codebooks.

\subsubsection{Task Settings}
We evaluate the proposed latent-space attacks in an image semantic communication scenario. Unless otherwise stated, all models are trained and tested on RGB images of resolution $256 \times 256$. We consider both reconstruction-oriented SemCom, where the receiver reconstructs the transmitted image, and task-oriented SemCom, where the receiver predicts downstream labels (e.g., object categories).

To decouple semantic vulnerabilities from physical-layer impairments, the channel between the semantic encoder and decoder is modeled as an error-free digital link. Thus, any degradation in reconstruction quality or task performance is solely caused by the semantic encoder–decoder and the injected latent-space perturbations.

\subsubsection{Attacker Settings}
We evaluate both the diffusion-based re-encoding attack (DiR) and the test-time adaptation latent manipulation attack (TTA-LM) under the same threat model. The attacker is assumed to be a Man-in-the-Middle adversary with full read–write access to the transmitted latent sequence, but without the ability to modify the encoder or decoder parameters.

\subsection{DiR Attack Evaluation}
\begin{figure*}[htbp]
    \centering
    \noindent\includegraphics[width=6in]{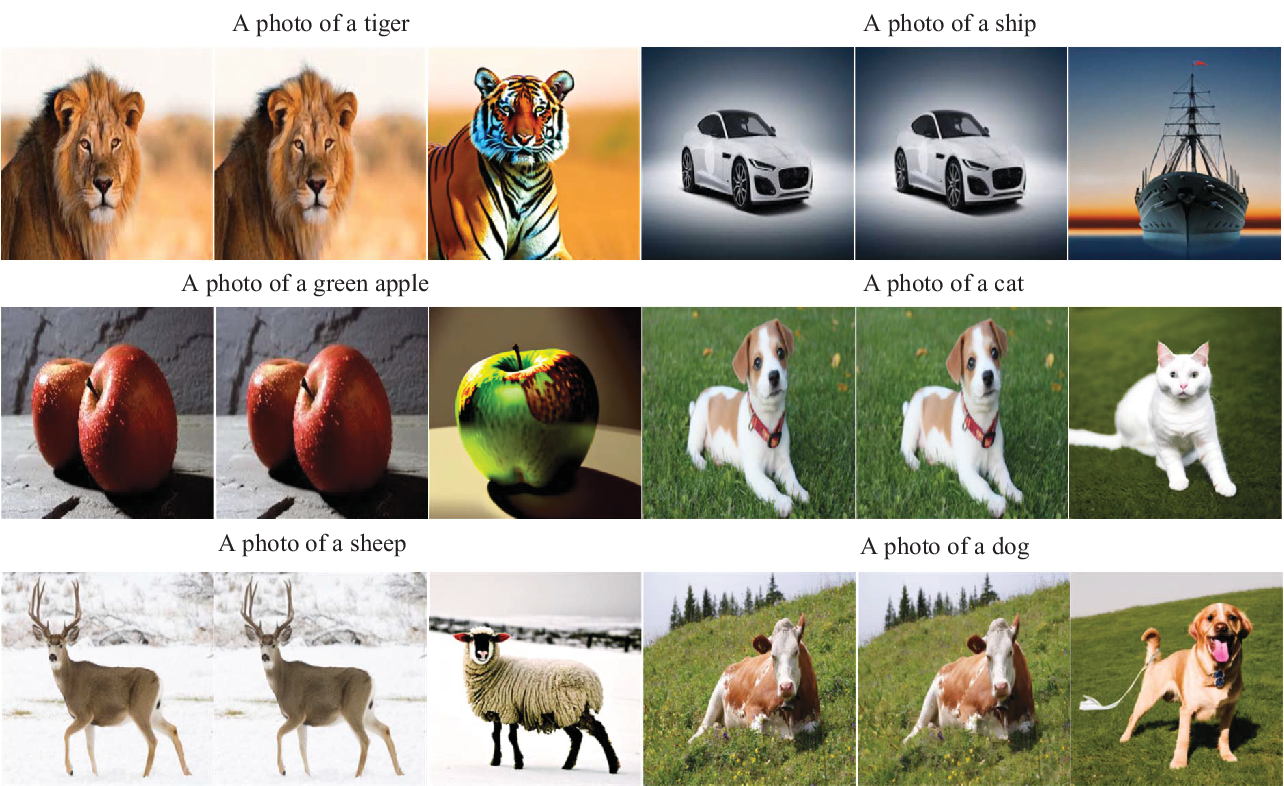}
    \caption{Reconstruction Results of DiR Attack}
    \label{dir_exp}
\end{figure*}

Fig.~\ref{dir_exp} illustrates reconstruction results obtained with the DiR under several target prompts, including “A photo of a tiger,” “A photo of a ship,” “A photo of a green apple,” “A photo of a cat,” “A photo of a sheep,” and “A photo of a dog.” In each case, DiR generates a latent representation whose decoded image closely matches the attacker-specified semantic description, yielding visually plausible tigers, ships, apples, cats, sheep, and dogs. The examples demonstrate that, by leveraging a latent diffusion model, DiR can impose strong and fine-grained semantic control over the reconstructed content, further confirming the practicality of latent-space semantic tampering.

\subsection{TTA-LM Attack Evaluation}
\begin{figure*}[htbp]
    \centering
    \noindent\includegraphics[width=6in]{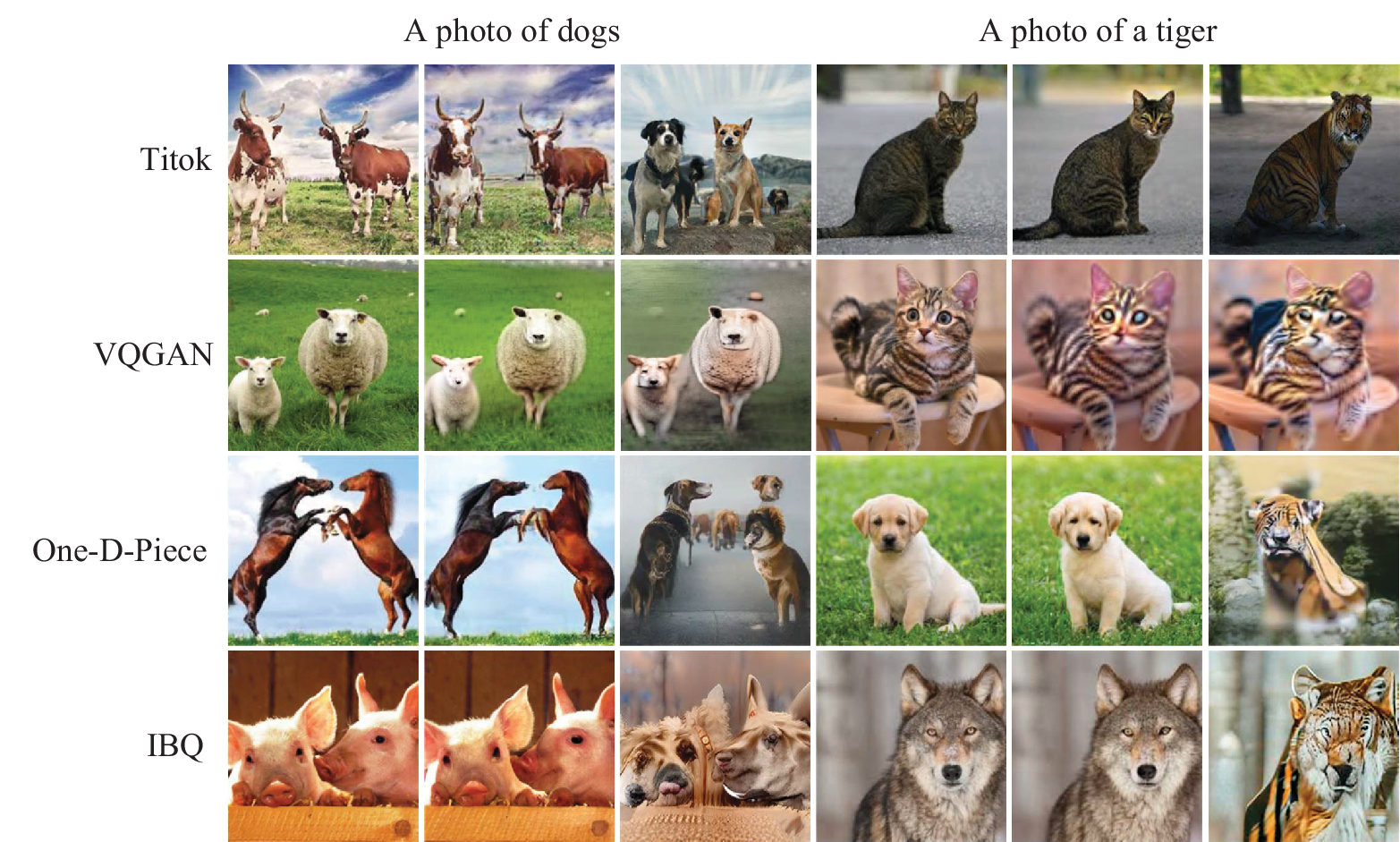}
    \caption{Reconstruction Results of TTA-LM Attack}
    \label{tto_exp}
\end{figure*}

Fig.~\ref{tto_exp} presents qualitative reconstruction results of the proposed TTA-LM attack on different semantic encoders. For each tokenizer (TiTok, VQGAN, One-D-Piece, and IBQ), we show decoded images after manipulating the intercepted latent representation toward two textual targets, “A photo of dogs” and “A photo of a tiger.” Across all four backbones, TTA-LM is able to steer the semantics of the reconstructed images toward the desired target concepts, producing dog-like or tiger-like appearances while maintaining realistic textures and backgrounds. This confirms that our gradient-based latent manipulation can consistently achieve targeted semantic shifts even when the underlying latent spaces and codebooks differ substantially.

\section{Conclusion}

In this paper, we revisited the security of deep learning–based semantic communication from the perspective of latent-space manipulation. We showed that, contrary to the common perception that semantic latent representations are difficult to control, an on-path adversary can efficiently and purposefully tamper with transmitted latents while keeping them statistically consistent with those produced by the legitimate encoder. To substantiate this finding, we proposed two complementary attack mechanisms: a Diffusion-based Re-encoding Attack (DiR), which leverages a latent diffusion model to synthesize attacker-specified semantics and re-encode them into valid latent representations, and a Test-Time Adaptation Latent Manipulation (TTA-LM) attack, which perturbs intercepted latents toward adversary-defined semantic objectives in a training-free and model-agnostic manner.

Extensive experiments on representative semantic encoders, including VQGAN, TiTok, One-D-Piece, and IBQ, demonstrated that both DiR and TTA-LM can induce targeted and perceptually plausible semantic changes in the decoded images, while preserving global latent-space statistics that make the attacks difficult to detect using simple distributional tests. These results reveal a fundamental and previously overlooked vulnerability in latent-space transmission for semantic communication systems.

Our study highlights the need to treat semantic integrity and latent-space authenticity as first-class security objectives in future SemCom designs. Promising directions for future work include developing lightweight cryptographic mechanisms and semantic consistency checks for protecting transmitted latents, extending latent-space attacks and defenses to multimodal and task-oriented semantic communication, and investigating robustness under more realistic channel conditions and system constraints.

\bibliographystyle{IEEEtran}
\bibliography{tto_attack}

@article{survey_1,
  title={Less data, more knowledge: Building next generation semantic communication networks},
  author={Chaccour, Christina and Saad, Walid and Debbah, M{\'e}rouane and Han, Zhu and Poor, H Vincent},
  journal={IEEE Communications Surveys \& Tutorials},
  year={2024},
  publisher={IEEE}
}

@article{secure_sc,
  title={Secure semantic communications: Fundamentals and challenges},
  author={Yang, Zhaohui and Chen, Mingzhe and Li, Gaolei and Yang, Yang and Zhang, Zhaoyang},
  journal={IEEE network},
  volume={38},
  number={6},
  pages={513--520},
  year={2024},
  publisher={IEEE}
}

@ARTICLE{uav_1,
  author={Sun, Geng and Xiao, Jian and Li, Jiahui and Wang, Jiacheng and Kang, Jiawen and Niyato, Dusit and Mao, Shiwen},
  journal={IEEE Transactions on Mobile Computing}, 
  title={Aerial Reliable Collaborative Communications for Terrestrial Mobile Users via Evolutionary Multi-Objective Deep Reinforcement Learning}, 
  year={2025},
  volume={24},
  number={7},
  pages={5731-5748},
  doi={10.1109/TMC.2025.3536093}}

@ARTICLE{uav_2,
  author={Li, Jiahui and Sun, Geng and Duan, Lingjie and Wu, Qingqing},
  journal={IEEE Transactions on Mobile Computing}, 
  title={Multi-Objective Optimization for UAV Swarm-Assisted IoT With Virtual Antenna Arrays}, 
  year={2024},
  volume={23},
  number={5},
  pages={4890-4907},
  doi={10.1109/TMC.2023.3298888}}

@ARTICLE{uav_3,
  author={Sun, Geng and Xie, Wenwen and Niyato, Dusit and Du, Hongyang and Kang, Jiawen and Wu, Jing and Sun, Sumei and Zhang, Ping},
  journal={IEEE Network}, 
  title={Generative AI for Advanced UAV Networking}, 
  year={2025},
  volume={39},
  number={4},
  pages={244-253},
  doi={10.1109/MNET.2024.3494862}}

@article{vqvae_sc,
  title={Robust semantic communications with masked VQ-VAE enabled codebook},
  author={Hu, Qiyu and Zhang, Guangyi and Qin, Zhijin and Cai, Yunlong and Yu, Guanding and Li, Geoffrey Ye},
  journal={IEEE Transactions on Wireless Communications},
  volume={22},
  number={12},
  pages={8707--8722},
  year={2023},
  publisher={IEEE}
}

@article{swinjscc,
  title={Swinjscc: Taming swin transformer for deep joint source-channel coding},
  author={Yang, Ke and Wang, Sixian and Dai, Jincheng and Qin, Xiaoqi and Niu, Kai and Zhang, Ping},
  journal={IEEE Transactions on Cognitive Communications and Networking},
  year={2024},
  publisher={IEEE}
}

@article{vit_sc,
  title={A robust image semantic communication system with multi-scale vision transformer},
  author={Peng, Xiang and Qin, Zhijin and Tao, Xiaoming and Lu, Jianhua and Letaief, Khaled B},
  journal={IEEE Journal on Selected Areas in Communications},
  year={2025},
  publisher={IEEE}
}

@article{RVQ_sc,
  title={Low-Bitrate High-Quality Digital Semantic Communication Based on RVQGAN},
  author={Chen, Xiaojiao and Wang, Jing and Huang, Jingxuan and Zeng, Ming and Zheng, Zhong and Fei, Zesong},
  journal={IEEE Internet of Things Journal},
  year={2025},
  publisher={IEEE}
}

@article{speech_swin_sc,
  title={Speech semantic communication based on swin transformer},
  author={Zhou, Ziliang and Zheng, Shilian and Chen, Jie and Zhao, Zhijin and Yang, Xiaoniu},
  journal={IEEE Transactions on Cognitive Communications and Networking},
  volume={10},
  number={3},
  pages={756--768},
  year={2023},
  publisher={IEEE}
}

@article{speech_sc,
  title={Semantic communication systems for speech transmission},
  author={Weng, Zhenzi and Qin, Zhijin},
  journal={IEEE Journal on Selected Areas in Communications},
  volume={39},
  number={8},
  pages={2434--2444},
  year={2021},
  publisher={IEEE}
}

@article{text_sc_1,
  title={A robust semantic text communication system},
  author={Peng, Xiang and Qin, Zhijin and Tao, Xiaoming and Lu, Jianhua and Hanzo, Lajos},
  journal={IEEE Transactions on Wireless Communications},
  volume={23},
  number={9},
  pages={11372--11385},
  year={2024},
  publisher={IEEE}
}

@article{text_sc_2,
  title={A GAN-based semantic communication for text without CSI},
  author={Mao, Jin and Xiong, Ke and Liu, Ming and Qin, Zhijin and Chen, Wei and Fan, Pingyi and Letaief, Khaled Ben},
  journal={IEEE Transactions on Wireless Communications},
  volume={23},
  number={10},
  pages={14498--14514},
  year={2024},
  publisher={IEEE}
}

@article{security_survey_1,
  title={A survey on semantic communication networks: Architecture, security, and privacy},
  author={Guo, Shaolong and Wang, Yuntao and Zhang, Ning and Su, Zhou and Luan, Tom H and Tian, Zhiyi and Shen, Xuemin},
  journal={IEEE Communications Surveys \& Tutorials},
  year={2024},
  publisher={IEEE}
}

@article{security_survey_2,
  title={Secure semantic communications: Challenges, approaches, and opportunities},
  author={Shen, Meng and Wang, Jing and Du, Hongyang and Niyato, Dusit and Tang, Xiangyun and Kang, Jiawen and Ding, Yaoling and Zhu, Liehuang},
  journal={IEEE Network},
  volume={38},
  number={4},
  pages={197--206},
  year={2023},
  publisher={IEEE}
}

@article{secure_sc_1,
  title={Is semantic communication secure? A tale of multi-domain adversarial attacks},
  author={Sagduyu, Yalin E and Erpek, Tugba and Ulukus, Sennur and Yener, Aylin},
  journal={IEEE Communications Magazine},
  volume={61},
  number={11},
  pages={50--55},
  year={2023},
  publisher={IEEE}
}

@article{secure_sc_2,
  title={Adversarial Attacks Against Shared Knowledge Interpretation in Semantic Communications},
  author={Hoang, Van-Tam and Nguyen, Van-Linh and Chang, Rong-Guey and Lin, Po-Ching and Hwang, Ren-Hung and Duong, Trung Q},
  journal={IEEE Transactions on Cognitive Communications and Networking},
  year={2025},
  publisher={IEEE}
}

@inproceedings{secure_sc_3,
  title={Backdoor attacks and defenses on semantic-symbol reconstruction in semantic communications},
  author={Zhou, Yuan and Hu, Rose Qingyang and Qian, Yi},
  booktitle={ICC 2024-IEEE International Conference on Communications},
  pages={734--739},
  year={2024},
  organization={IEEE}
}

@inproceedings{secure_sc_4,
  title={Stealthy Backdoor Attacks on Semantic Symbols in Semantic Communications},
  author={Zhou, Yuan and Hu, Rose Qingyang and Qian, Yi},
  booktitle={GLOBECOM 2024-2024 IEEE Global Communications Conference},
  pages={4975--4981},
  year={2024},
  organization={IEEE}
}

@inproceedings{secure_sc_5,
  title={Imperceptible and Targeted Physical Attacks on Deep Learning-Based Speech Semantic Communications},
  author={Hua, Yuhao and Xu, Yang and Lyu, Chen and Liu, Jia and Shen, Yulong and Yang, Weidong and Shiratori, Norio},
  booktitle={2025 IEEE Wireless Communications and Networking Conference (WCNC)},
  pages={1--6},
  year={2025},
  organization={IEEE}
}

@inproceedings{secure_sc_6,
  title={SemPerGe: Unveiling Text-Based Adversarial Attacks on Semantic Communication},
  author={Anjum, Afia and Mitra, Arkajyoti and Agbaje, Paul and Alam, Md Ahanaful and Roy, Debashri and Parwez, Md Salik and Olufowobi, Hebeeb},
  booktitle={2025 IEEE Conference on Communications and Network Security (CNS)},
  pages={1--9},
  year={2025},
  organization={IEEE}
}

@inproceedings{secure_sc_7,
  title={The model inversion eavesdropping attack in semantic communication systems},
  author={Chen, Yuhao and Yang, Qianqian and Shi, Zhiguo and Chen, Jiming},
  booktitle={GLOBECOM 2023-2023 IEEE Global Communications Conference},
  pages={5171--5177},
  year={2023},
  organization={IEEE}
}

@article{secure_sc_pls,
  title={Secure semantic communications: From perspective of physical layer security},
  author={Li, Yongkang and Shi, Zheng and Hu, Han and Fu, Yaru and Wang, Hong and Lei, Hongjiang},
  journal={IEEE Communications Letters},
  year={2024},
  publisher={IEEE}
}

@article{secure_sc_blackbox,
  title={Secure semantic communication model for black-box attack challenge under metaverse},
  author={Li, Chang and Zeng, Liang and Huang, Xin and Miao, Xiaqing and Wang, Shuai},
  journal={IEEE Wireless Communications},
  volume={30},
  number={4},
  pages={56--62},
  year={2023},
  publisher={IEEE}
}

@inproceedings{secure_sc_genai,
  title={Generative Al-aided joint training-free secure semantic communications via multi-modal prompts},
  author={Du, Hongyang and Liu, Guangyuan and Niyato, Dusit and Zhang, Jiayi and Kang, Jiawen and Xiong, Zehui and Ai, Bo and Kim, Dong In},
  booktitle={ICASSP 2024-2024 IEEE International Conference on Acoustics, Speech and Signal Processing (ICASSP)},
  pages={12896--12900},
  year={2024},
  organization={IEEE}
}

@inproceedings{secure_sc_img,
  title={Secure semantic communication for image transmission in the presence of eavesdroppers},
  author={Tang, Shunpu and Liu, Chen and Yang, Qianqian and He, Shibo and Niyato, Dusit},
  booktitle={GLOBECOM 2024-2024 IEEE Global Communications Conference},
  pages={2172--2177},
  year={2024},
  organization={IEEE}
}

@article{secure_sc_blockchain,
  title={Blockchain-aided secure semantic communication for AI-generated content in metaverse},
  author={Lin, Yijing and Du, Hongyang and Niyato, Dusit and Nie, Jiangtian and Zhang, Jiayi and Cheng, Yanyu and Yang, Zhaohui},
  journal={IEEE Open Journal of the Computer Society},
  volume={4},
  pages={72--83},
  year={2023},
  publisher={IEEE}
}

@inproceedings{ldm,
  title={High-resolution image synthesis with latent diffusion models},
  author={Rombach, Robin and Blattmann, Andreas and Lorenz, Dominik and Esser, Patrick and Ommer, Bj{\"o}rn},
  booktitle={Proceedings of the IEEE/CVF conference on computer vision and pattern recognition},
  pages={10684--10695},
  year={2022}
}

@inproceedings{hc_tokenizer,
  title={Highly Compressed Tokenizer Can Generate Without Training},
  author={Beyer, Lukas Lao and Li, Tianhong and Chen, Xinlei and Karaman, Sertac and He, Kaiming},
  booktitle={Forty-second International Conference on Machine Learning}
}

@inproceedings{recon_gen,
  title={Reconstruction vs. generation: Taming optimization dilemma in latent diffusion models},
  author={Yao, Jingfeng and Yang, Bin and Wang, Xinggang},
  booktitle={Proceedings of the Computer Vision and Pattern Recognition Conference},
  pages={15703--15712},
  year={2025}
}

@inproceedings{latentpaint,
  title={Latentpaint: Image inpainting in latent space with diffusion models},
  author={Corneanu, Ciprian and Gadde, Raghudeep and Martinez, Aleix M},
  booktitle={Proceedings of the IEEE/CVF winter conference on applications of computer vision},
  pages={4334--4343},
  year={2024}
}

@inproceedings{sdgan,
  title={SDGAN: disentangling semantic manipulation for facial attribute editing},
  author={Huang, Wenmin and Luo, Weiqi and Huang, Jiwu and Cao, Xiaochun},
  booktitle={Proceedings of the AAAI conference on artificial intelligence},
  volume={38},
  number={3},
  pages={2374--2381},
  year={2024}
}

@article{titok,
  title={An image is worth 32 tokens for reconstruction and generation},
  author={Yu, Qihang and Weber, Mark and Deng, Xueqing and Shen, Xiaohui and Cremers, Daniel and Chen, Liang-Chieh},
  journal={Advances in Neural Information Processing Systems},
  volume={37},
  pages={128940--128966},
  year={2024}
}

@inproceedings{vqgan,
  title={Taming transformers for high-resolution image synthesis},
  author={Esser, Patrick and Rombach, Robin and Ommer, Bjorn},
  booktitle={Proceedings of the IEEE/CVF conference on computer vision and pattern recognition},
  pages={12873--12883},
  year={2021}
}

@article{one_d,
  title={One-d-piece: Image tokenizer meets quality-controllable compression},
  author={Miwa, Keita and Sasaki, Kento and Arai, Hidehisa and Takahashi, Tsubasa and Yamaguchi, Yu},
  journal={arXiv preprint arXiv:2501.10064},
  year={2025}
}

@inproceedings{ibq,
  title={Scalable image tokenization with index backpropagation quantization},
  author={Shi, Fengyuan and Luo, Zhuoyan and Ge, Yixiao and Yang, Yujiu and Shan, Ying and Wang, Limin},
  booktitle={Proceedings of the IEEE/CVF International Conference on Computer Vision},
  pages={16037--16046},
  year={2025}
}

\end{document}